\documentclass[twocolumn]{aastex61}
\usepackage{natbib}

\newcommand{\logg}{\ensuremath{\log g}}

\newcommand\aastex{AAS\TeX}

\received{\today}

\shorttitle{\aastex\ A carbon rich ultra metal-poor dwarf star}
\shortauthors{Aguado et al.}

\begin{document}
   \title{J0815+4729: A chemically primitive dwarf star in the Galactic Halo observed with Gran Telescopio Canarias.\footnote{
 Based on observations made with the Gran Telescopio Canarias (GTC), 
 installed in the Spanish Observatorio del Roque de los Muchachos 
 of the Instituto de Astrof\'{\i}sica de Canarias, on the island of La Palma.
 Program ID GTC90-15B and the Discretionary Director Time GTC03-16ADDT.}
 }

\correspondingauthor{David~S. Aguado}
\email{aguado@iac.es}

\author{David~S. Aguado}
\affil{Instituto de Astrof\'{\i}sica de Canarias,
             V\'{\i}a L\'actea, 38205 La Laguna, Tenerife, Spain\\}
 \affiliation{Universidad de La Laguna, Departamento de Astrof\'{\i}sica, 
             38206 La Laguna, Tenerife, Spain \\}             

\author{Jonay~I. Gonz\'alez Hern\'andez}
\affil{Instituto de Astrof\'{\i}sica de Canarias,
              V\'{\i}a L\'actea, 38205 La Laguna, Tenerife, Spain\\}
 \affiliation{Universidad de La Laguna, Departamento de Astrof\'{\i}sica, 
             38206 La Laguna, Tenerife, Spain \\}             

\author{ Carlos Allende Prieto}
\affil{Instituto de Astrof\'{\i}sica de Canarias,
              V\'{\i}a L\'actea, 38205 La Laguna, Tenerife, Spain\\}
 \affiliation{Universidad de La Laguna, Departamento de Astrof\'{\i}sica, 
             38206 La Laguna, Tenerife, Spain \\}             

\author{Rafael Rebolo}
\affil{Instituto de Astrof\'{\i}sica de Canarias,
              V\'{\i}a L\'actea, 38205 La Laguna, Tenerife, Spain\\}
 \affiliation{Universidad de La Laguna, Departamento de Astrof\'{\i}sica, 
             38206 La Laguna, Tenerife, Spain \\}             

\affiliation{ Consejo Superior de Investigaciones Cient\'{\i}ficas, 28006 Madrid, Spain\\}




\begin{abstract}
We report the discovery of the carbon-rich ultra metal-poor unevolved star J0815+4729.  
This dwarf star was selected from SDSS/BOSS as a metal-poor candidate and follow-up spectroscopic observations at medium-resolution  were obtained with
ISIS at William Herschel Telescope and OSIRIS at Gran Telescopio de Canarias.
We use the FERRE code to derive the main stellar parameters, $T_{\rm eff}=$6215$\pm 82$\,K, and \logg=4.7$\pm0.5$,  an upper limit to the metallicity of [Fe/H]$\leq-5.8$, and a carbon abundance of [C/Fe]$\geq+5.0$, while $[\alpha/\rm Fe]$=0.4 is assumed. The metallicity upper limit is based on the \ion{Ca}{2} K line, that at the resolving power of  the OSIRIS spectrograph cannot be resolved from possible interstellar calcium. The star could be the most iron-poor unevolved star known, and is also amongst the ones with the largest overabundances of carbon. High-resolution spectroscopy of J0815+4729 will certainly help to derive other important element abundances, possibly providing new fundamental constraints on the early stages of the Universe, the formation of the first stars and the properties of the first supernovae.

\end{abstract}




\section{Introduction} \label{intro}
The existence of  surviving population III stars is still under debate. The issue is intimately linked to the minimum mass at which a star can form at zero metallicity. Second generation stars are formed from matter polluted by the first supernovae; their chemical composition reflects the yields from the first massive stars. We can study the early chemical evolution of the Universe through the analysis of those second-generation low-mass stars.  Stellar archeology has a deep impact in several fields of modern astrophysics, from stellar formation and evolution, to near-field cosmology. To make progress, it is necessary to identify larger samples of primitive stars and derive their chemical abundances. 

The number of known extremely metal-poor stars has dramatically increased since the 80's. G64-12 with [Fe/H]$=-3.2$\footnote{We use the bracket notation for reporting
chemical abundances: [a/b]$ = \log \left( \frac{\rm N(a)}{\rm N(b)}\right) - \log
\left( \frac{\rm N(a)}{\rm N(b)}\right)_{\odot}$,
where $\rm N$(x) represents number density of nuclei of the element x.}  was discovery by \citet{car81}, but these authors originally reported a metallicity of [Fe/H]$=-3.52$. Nowadays, a few hundred stars are cataloged as having [Fe/H]$<-3.0$ (see e.g. \citet{bee05,aoki06I,caff13I,norris13I,yong13II,pla15,agu16} and references therein). 

The first ultra metal-poor star ([Fe/H]$<-4.0$)  was spectroscopically studied by \citet{bes84}: CD --38$^{\circ}$ 245,  with [Fe/H]$=-4.5$. Recent studies by \citet{yong13II} derive  a metallicity of [Fe/H]$=-4.15$ for this star. Nearly thirty ultra metal-poor stars are already known (see e.g. \cite{boni12,roe14,yong13II,han14,boni15,pla15,alle15,agu17I} and references therein).
\citet{chris02,chris04} reported from Hamburg/ESO survey the discovery of HE 0107--5240 with a derived metallicity of [Fe/H]$=-5.39$ (LTE value). Since then, four more metal-poor stars below  [Fe/H]$=-5$ have been detected: HE 1327--2326 with [Fe/H]$=-5.65$ \citep{fre05,aoki06I}; J1035+0641 with  [Fe/H]$<-5.07$ \citep{boni15} from the Sloan Sky Digital Survey (SDSS) \citep{yor00}; SMSS J01313--6708 with no iron features detected and  [Fe/H]$<-7.20$ \citep{kel14}, and finally J1029+1729  \citep{caff11} with [Fe/H]$=-5.0$ but no measurable carbon abundance ([C/Fe]$<+0.7$).

With the exception of J1029+1729, all known metal-poor stars at [Fe/H]$<-4.5$,  including J0815+4729, are carbon-enhanced metal-poor stars (CEMP).  The observed increase in the frequency of CEMP stars at the lowest metallicities (see, e.g. \citealt{coh05,pla14}) could be explained by the fall-back mechanism in core-collapse supernovae (SN) \citep{ume03} which is an easy way to get high C and low Fe
out into the interstellar medium in the early universe. This effect is most relevant for zero-metallicity supernova progenitors (e.g. \citealt{lim03}). In addition, \citet{coo14} suggest the importance of the ability of the host minihalos to retain their gas reservoir in order to explain the chemical composition observed in second-generation stars.

 The calcium K resonance line is the strongest detectable metallic absorption line in the visible spectrum~\citep[see e.g.][]{bee92}. This line allows us to derive a metallicity estimate assuming $[\alpha/\rm Fe]$=0.4, even when iron lines are not detected, using medium resolution spectroscopy \citep{agu17I,agu17II}.
 In this work we report the discovery of J0815+4729, a new carbon-rich ultra metal-poor dwarf star with [Fe/H]$\leq-5.8$. The target identification and observations are explained in Section \ref{obs}.
The determination of atmospheric parameters is described in Section \ref{anali} together with a complete analysis of two comparison well-known metal-poor stars, G64-12 and J1313-0019. The discussion and the conclusions are given in Section \ref{conclusion}.
\begin{figure}
\begin{center}
{\includegraphics[width=90 mm, angle=0]{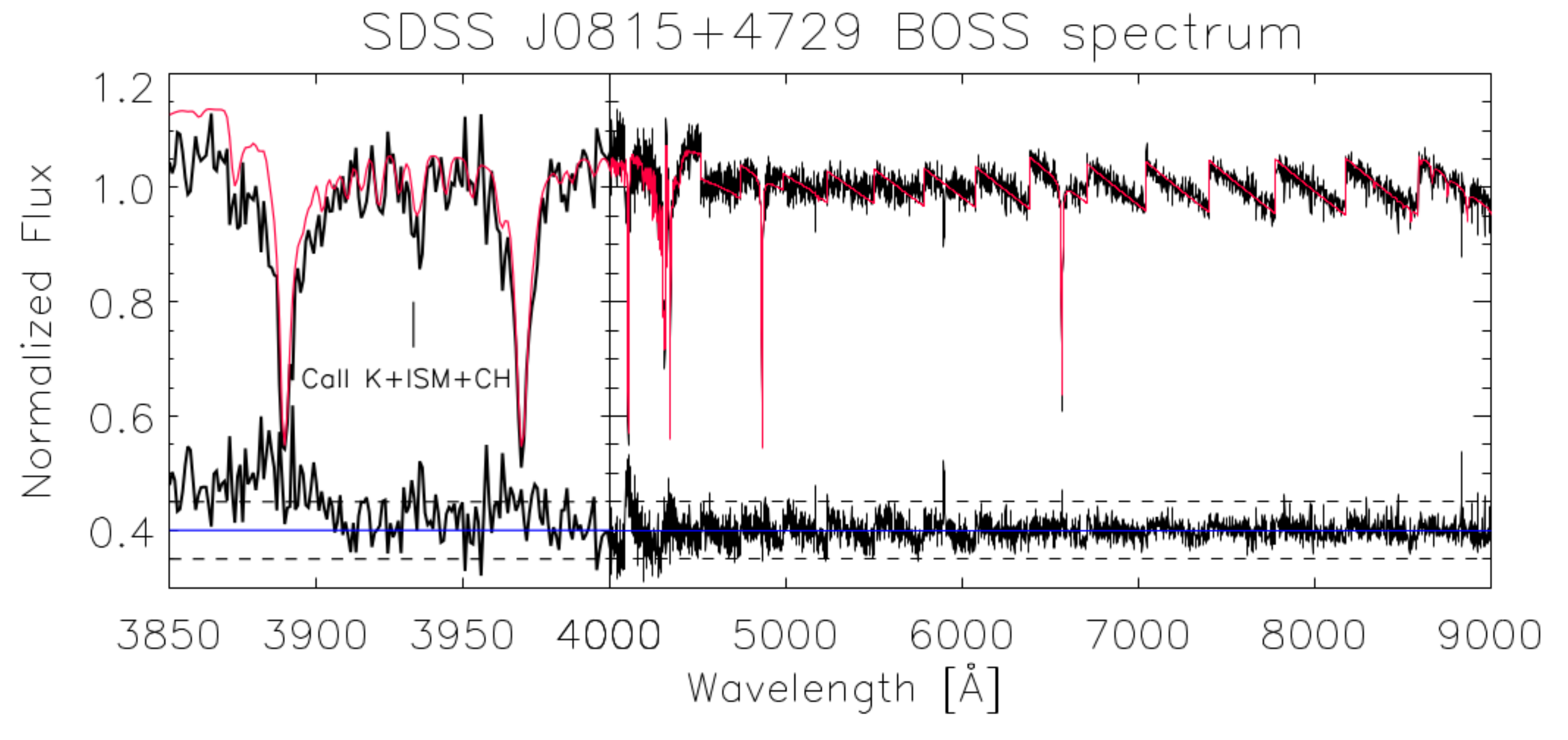}}
\end{center}
\caption{BOSS spectrum of J0815+4729 (black line) and the best fit obtained with FERRE (red line), after dividing the spectra in  segments  of about $\sim$300-400\AA\, 
which are normalized by their average fluxes. The residuals are shown in the lower panel.}
\label{boss}
\end{figure}

\section{Target selection and Observations} \label{obs}
 We analyzed more than 
2.5 million low-resolution spectra from the Sloan  Extension for Galactic Understanding and Exploration (SEGUE, \citealt{yan09}),
the Baryonic Oscillations Spectroscopic Survey (BOSS, \citealt{eis11,daw13}) and the large Sky Area 
Multi-Object Fiber Spectroscopic Telescope (LAMOST, \citealt{deng12}).
We derived a first set of stellar parameters: effective temperature, surface gravity, metallicity and
carbon abundance (see Section \ref{anali}). This methodology allowed us to built a catalogue of $\sim100$ metal-poor candidates to observe using medium-resolution follow-up spectroscopy with the Intermediate dispersion Spectrograph and Imaging System (ISIS)
at the 4.2m William Herschel Telescope (WHT) \citep{agu16,agu17I} in the {\it Observatorio del Roque de los Muchachos} (La Palma, Spain).
Our first analysis of the BOSS spectrum of J0815+4729 (R.A.$=08^{h} 15^{m} 54^{s}.25$, DEC.=$+47^{0} 29^{'} 47^{''}.85$ (J2000), mag(g)=17.06$\pm 0.01$ and $v_{rad}= -118\pm 5$\,km s$^{-1}$.) with FERRE \footnote{{\tt FER\reflectbox{R}E} is available 
from http://github.com/callendeprieto/ferre} provided the parameters $T_{\rm eff}=6365$\,K,
$\logg$=4.9, $\left[{\rm Fe/H}\right]$=-4.6  and $\left[{\rm C/Fe}\right]$=+3.9, with FERRE internal uncertainties of 19\,K; 0.06; 0.10\,dex; 0.09\,dex, respectively (see Figure \ref{boss}).  J0815+4729 was selected as a promising candidate to be extremely metal-poor as inferred in the FERRE analysis from the weak absorption shown in the region of the calcium K line. Further details about the selection candidates could be found in \citet{agu17I}.

\begin{figure}
\begin{center}
{\includegraphics[width=90 mm]{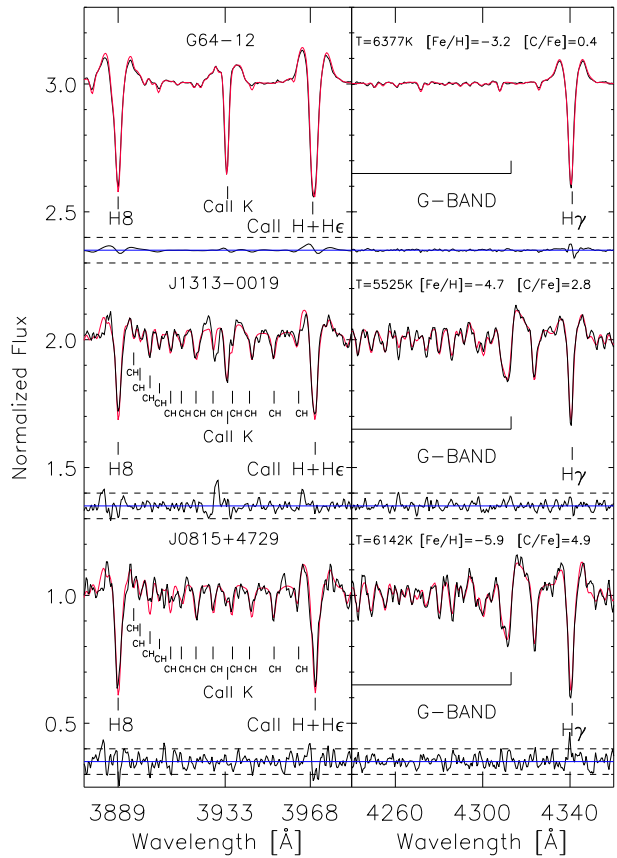}}
\end{center}
\caption{ISIS/WHT spectra of the G-band and \ion{Ca}{2} lines of the stars J0815+4729, J1313-0019 and G64-12
(black line) and the best fit calculated with FERRE (red line). Both the observed and synthetic spectra have been normalized with a running-mean filter with a width of 30 pixels. The normalized spectra are vertically shifted for clarity. The residuals  (diference
between the observed spectrum and the best fit) are displayed under each spectrum, and the dashed lines  correspond to differences of  $+5\%$ and $-5\%$.
 The effective temperature, metallicity and carbon abundance for each spectrum are also displayed.}
\label{isis}
\end{figure}

Some obvious imperfections in the normalization of the BOSS spectrum probably caused by issues with the  flux calibration together with significant noise in the vicinity of the calcium K line warranted  a spectrum of higher S/N.
The ISIS observations were obtained between Dec 31, 2014 - Jan 2, 2015, program C103.
The set-up adopted the R600B and R600R gratings, the GG495 filter in the red arm and the 
default dichroic (5300 \AA). The mean resolving power using a  1\farcs slit was 
$R \sim 2400$ in the blue arm and $R \sim 5200$ in the red arm.
 Six individual exposures of 1800\,s were taken to avoid a significant degradation of the
data due to cosmic rays and a coadded spectrum 
of S/N$ \sim 55$ was obtained. In addition,  two other well-studied metal-poor stars,
G64-12 and J1313-0019 were observed for calibrating purposes (See Fig. \ref{isis}) and 
details of the observations are provided in \citet{agu17I}. 

After the analysis of the ISIS spectrum confirmed a very low metallicity we decided to obtain new spectrum with a much higher signal-to-noise ratio using the 10.4m Gran Telescopio Canarias  (GTC) telescope equipped with the Optical System for Imaging and low-intermediate-Resolution
Integrated Spectroscopy (OSIRIS) instrument. This allowed us to secure a S/N$>$200 spectrum in a reasonable amount of time.

The observations were scheduled in service mode during the Feb 13, 2016 night as part of the GTC90-15B program and Mar 27, 2016 in GTC03-16ADDT program. The sky requirements were in both 
cases a seeing with a  FWHM$<1.2^{''}$ and gray conditions. We selected the R2500U grism of OSIRIS 
and a 1\farcs slit, providing a spectral range 3600-4500\,\AA\ with a resolving power
R$ \sim$2500. Ten exposures of 1584\,s were taken with identical set-up and similar sky conditions.
For both the ISIS and the OSIRIS data reduction (bias subtraction, flat-fielding and wavelength 
calibration, using CuNe $+$ CuAr lamps) we adopted the \emph{onespec} 
package in IRAF\footnote{IRAF is distributed by the National Optical Astronomy Observatory, 
which is operated by the Association of Universities for Research in Astronomy 
(AURA) under cooperative agreement with the National Science Foundation} \citep{tod93}. For further details see \citet{agu16,agu17I}.
\section{Analysis}\label{anali}
The analysis of J0815+4729 was carried out as explained in \citet{agu17I,agu17II}.
We used the grid of synthetic spectra \citep{agu17I} already available from CDS\footnote{{The model spectra are only available at the Centre de Donn\'ees astronomiques de Strasbourg (CDS) via anonymous
ftp to cdsarc.u-strasbg.fr (130.79.128.5) or via
http://cdsarc.u-strasbg.fr/viz-bin/qcat?J/A+A/605/A40}}. 
The grid was computed with the ASS$\epsilon$T code \citep{koe08} and uses the Barklem 
theory for self-broadening of the Balmer lines and the stark-broadened profiles from \citet{ste10}. Model atmospheres were computed with ATLAS9 \citep{kur79} as described by \citet{mez12}.  

We assumed $[\alpha/\rm Fe]$=0.4 as a canonical value for the Galactic halo \citep{yong13II}.  The most metal-poor stars known ($[\rm Fe/\rm H]\lesssim -4.5$) typically show 1D-LTE abundance ratios [CaII/Fe]  $\sim 0.25-0.4$~dex \citet{chris04,han14,alle15,fre15,boni15}. \citet{caff12I} estimate [\ion{Ca}{2}/Fe]$\sim0.13$, while \citet{fre08} get [\ion{Ca}{2}/Fe]~$\sim0.7$, for their stars.
On the other hand, \citet{emm15}  found  that $[\rm Ca/\rm Fe]$ depends on  $[\rm Fe/\rm H]$ and distance $r$ to the Galactic center, with higher values at lower  $[\rm Fe/\rm H]$ and slightly increasing at distances $r>20$~kpc, in the range 0.4-0.6~dex. Following \citet[][and references therein]{emm15},  we estimate a distance to the star of $\sim 2.3$~kpc and a Galactocentric distance of $\sim10$ kpc. Therefore, it seems reasonable to assume for J0815+4729 that $[\rm Ca/\rm Fe] \sim 0.4$. 
Synthetic spectra for metallicities below $\mbox{[Fe/H]}=-5$  were computed using model atmospheres with $\mbox{[Fe/H]}=-5$.
The limits of the grid of synthetic spectra adopted in our analysis are $4750\,\rm K\leq T_{eff}\leq7000\,\rm K$, $-6\leq[\rm Fe/H]\leq-2$; $+1 \leq[\rm C/Fe]\leq +5$, 
and $1.0\leq \logg \leq5.0$  and a microturbulence $\xi$ of 2\,km s$^{-1}$ was adopted.
The FERRE code is able to derive simultaneously the stellar parameters, including metallicity, 
and the carbon abundance. The observed and synthetic spectra are normalized with a  running-mean filter with a width 30 pixels (see \citet{agu17I} for further details).
\subsection{ISIS Analysis}\label{analisis}
The ISIS spectrum shows a myriad of CH transitions. These must be properly modeled to perform a thorough study of the \ion{Ca}{2} K line at 3933\,\AA\, which is also blended with a interstellar medium (ISM) contribution. Most halo stars with metallicities [Fe/H]$<-4.5$ show interstellar calcium absorption. That fact could be partially explained due most of they are at large distances. A Gaussian profile was adopted to model the calcium ISM contribution constructing a 
 grid of absorption features running from 5\% to 30\% with a step of 1\%, and relative velocities ($\Delta v$) from +70\,km s$^{-1}$ to +110\,km s$^{-1}$ with a step of 5\,km s$^{-1}$, until we got the best $\chi^{2}$, and the resulting spectrum was reanalyzed with FERRE. The minimum $\chi^{2}$ corresponds to a 28\% the ISM contribution to the stellar \ion{Ca}{2} K line at $\Delta v=+100$\,km/s.
\begin{figure*}
\begin{center}
{\includegraphics[width=180 mm, angle=180]{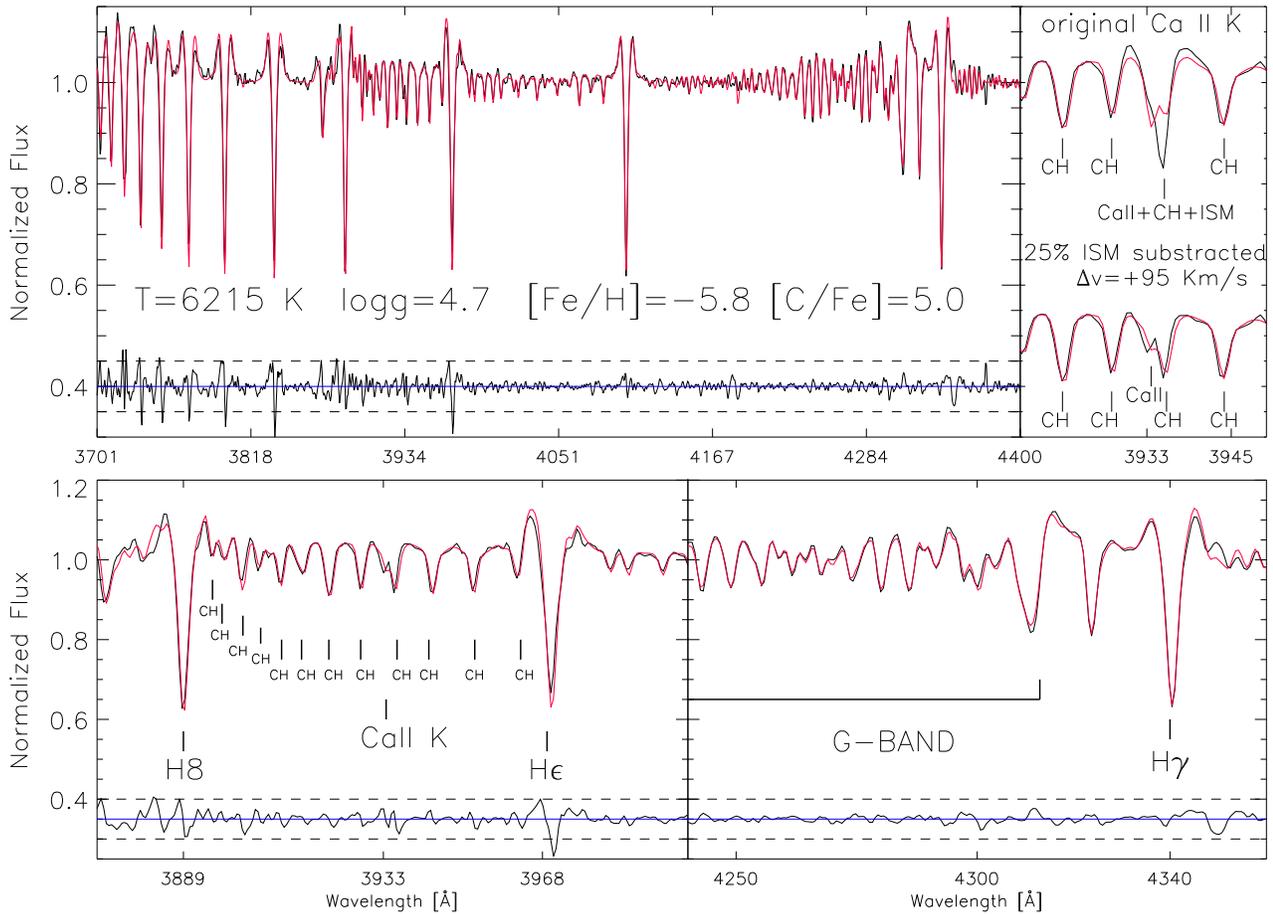}}
\end{center}
\caption{OSIRIS medium-resolution spectrum of J0815+4729 (black line) and the best fit obtained with FERRE (red line). The residuals (difference between the observed spectrum and the best
fit) are shown underneath each of the spectra, together with the $\pm$5\%reference lines.  The upper-right panel shows the original spectrum in the vicinity of the Ca II K line (top spectrum), and the same area after subtracting the ISM contribution 
(bottom).}

\label{cemp}
\end{figure*}

Figure \ref{isis} shows the ISIS spectrum of J0815+4729 together with other two well-known metal-poor stars, J1313-0019 and G64-12, also analyzed in \citet{agu17I}.  The right-hand side of the panel shows the G-band area. The stars J1313-0019 and J0815+4729 show a strong G-band whereas G64-12, according with \citet{agu17I}, is not carbon-enhanced. The CH features in J0815+4729 are indeed stronger than in J1313-0019 despite the latter is almost 600\,K cooler.
 The left-hand panels show the \ion{Ca}{2} K line for these three stars. From the ISIS analysis we arrive at $T_{\rm eff}=6142 \pm 118$\,K; $\logg=4.7 \pm0.6$; $\left[{\rm C/Fe}\right]=4.9\pm0.2$\,dex; and with the \ion{Ca}{2} K ISM feature at $\Delta v=+100$\,km/s,  $\left[{\rm Fe/H}\right]=-5.9\pm 0.2$\,dex  for J0815+4729. The derived radial velocity of the star is $v_{rad}= -83\pm 38$\,km s$^{-1}$.

\subsection{OSIRIS analysis}\label{analosiris}
 The much higher S/N of the OSIRIS spectrum allows us to better model the ISM contribution. 
The methodology followed is the same explained in \ref{analisis} and led to a more reliable value of 25\% absorption at  $\Delta v=+95$\,km/s, corresponding to an equivalent width of $\sim$30\,m\AA.
Figure \ref{cemp} shows the entire OSIRIS spectrum of J0815+4729 (upper-left panel) and the best fit derived with FERRE. 
 Without subtracting the ISM contribution (See Fig. \ref{cemp}, upper-right panel) FERRE is not able to model the blended line, thus overestimating the metallicity to be   $\left[{\rm Fe/H}\right]=-5.0$.
The \ion{Ca}{2} H\&K and G-band spectral regions are displayed in the bottom panel. 
The radial velocity obtained  from the OSIRIS spectrum by a cross correlation is $v_{rad}= -95\pm 23$\,km s$^{-1}$. The radial velocity values from the three spectra suggest no variation.  
The goodness of fit for the OSIRIS spectrum is remarkably good with a $\chi^{2}=1.1$ , leading to the values
T$_{\rm eff}=6215\pm 82$\,K, \logg=4.7$\pm 0.5$, [Fe/H]$\leq-5.8$\,dex and a carbon abundance of [C/Fe]$\geq+5.0$.
 At this extremely low metallicity, the majority of the information on metallicity is coming from the \ion{Ca}{2} K line, which provides an upper limit of [Ca/H]$<$-5.4 for J0815+4729. Since the structure of the ISM contribution could be more complicated than a separated single Gaussian component, and there could be more ISM components closely blended with the \ion{Ca}{2} K stellar feature (see e.g. \citealt{fre05,caff11,agu17I}), we consider the calcium abundance and, consequently, the metallicity value derived from the \ion{Ca}{2} K line as an upper limit.  

Small traces of Fe I lines can be observed in the blue part of the OSIRIS spectrum (3815-3860\,\AA). Unfortunately, these features are too weak to derive an iron abundance at the OSIRIS resolution. However, we compare in Figure 4 (upper panel) models with different iron abundances and the 
original spectrum.  It is clear that J0815+4729 is an iron-poor star with, at least, [Fe/H]~$ < -4.5$.
 J0815+4729 is the second most carbon-rich metal-poor star after SMSS J0313$-$6708 \citep{kel14}, 
which shows [C/Fe]$>$5.4 with a metallicity of [Fe/H]$<$-7.2 derived from the non-detection of the strongest Fe
lines and [Ca/H]$=$-7.2 derived from the \ion{Ca}{2} K line.
 Our analysis has been performed assuming a microturbulence $\xi=2.0$\,km s$^{-1}$ instead of the more suitable value for dwarf stars of $\xi \sim 1.5$\,km s$^{-1}$ \citep[see e.g.][]{cohe04,bar05}. We have done a simple test to evaluate the impact of this assumption. We analyzed with FERRE both synthetic spectra computed with the same stellar parameters of J0815-4729 but with two different $\xi$ values, 2.0 and 1.0\,km s$^{-1}$. The variation in metallicity is small (0.06\,dex) and the direction of higher metallicity with $\xi=1.0$\,km s$^{-1}$, as expected. This result is consistent with the derived upper limit [Fe/H]$\leq-5.8$.
\begin{figure}
\begin{center}\label{fam}
{\includegraphics[angle=180,width=90 mm]{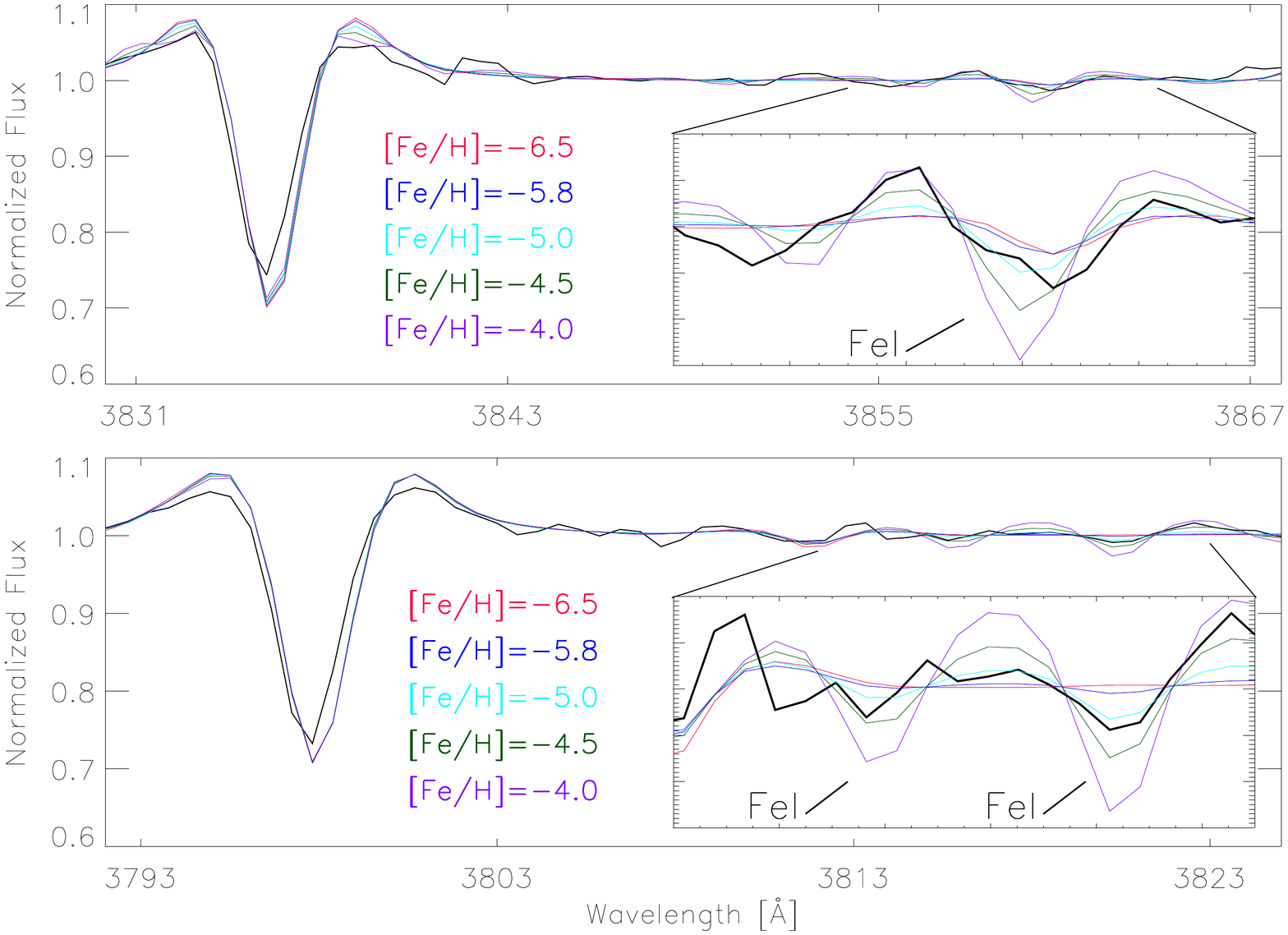}}
{\includegraphics[width=90 mm]{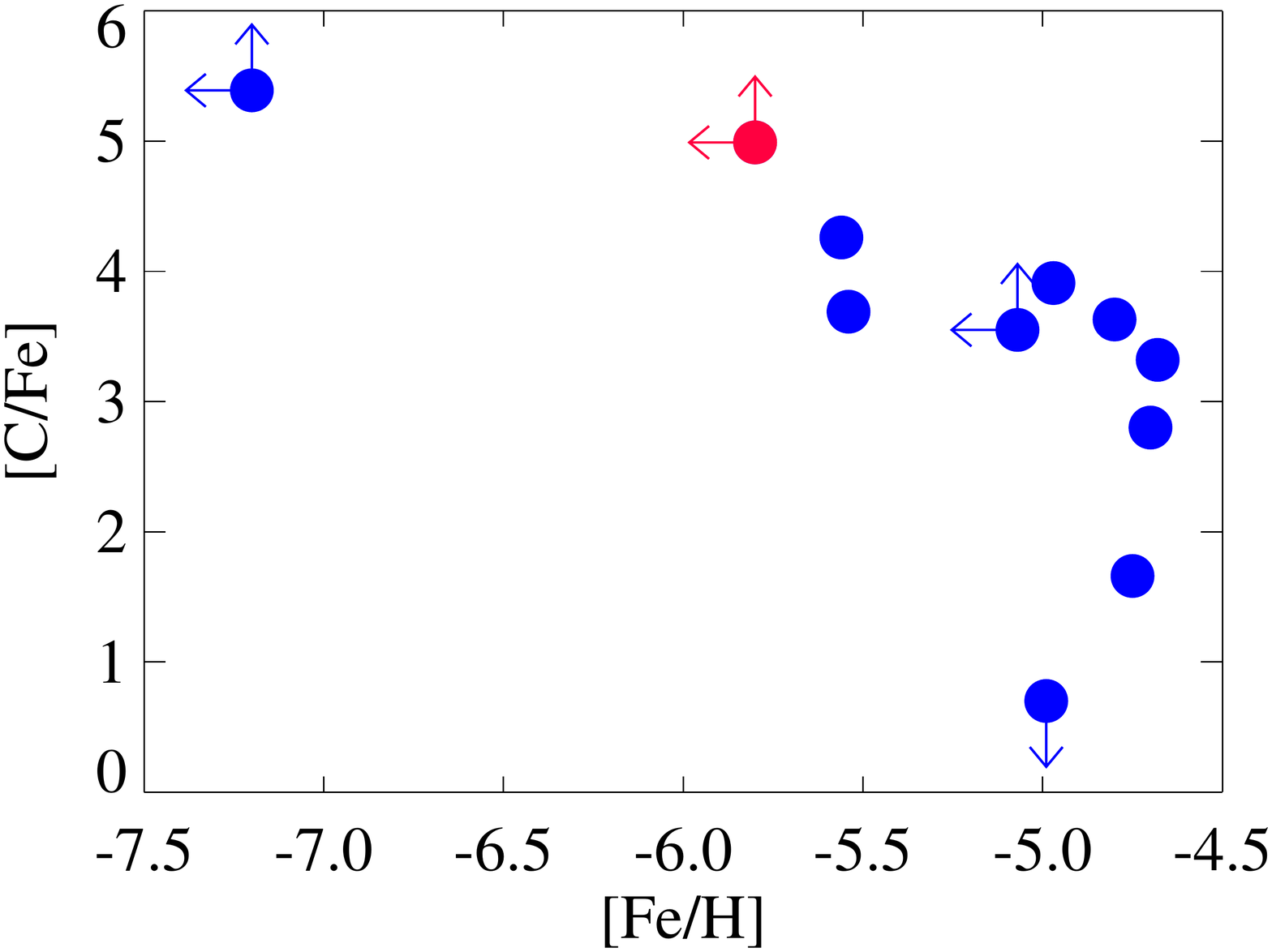}}
\end{center}
\caption{Upper panel: A detail of the OSIRIS J0815+4729 spectrum (black line) in the region covering strongest \ion{Fe}{1} transitions (3815, 3820 and 3859\,\AA). Five synthetic spectra with different metallicities are also shown.
Bottom panel: Carbon-to-iron ratio vs. iron abundance in the ten stars in the [Fe/H]$<$-4.5 regime already in the literature: \citet{chris04,fre05,nor07,caff11,han14,kel14,alle15,boni15}.
J0815+4729 is marked in red.}
\end{figure}
\section{Discussion and conclusions}\label{conclusion}
J0815+4729 is a main-sequence star  (T$_{\rm eff}=6215\pm 82$\,K, \logg=4.7$\pm 0.5$) with a metallicity of [Fe/H]$\leq-5.8$\,dex.  Finding unevolved stars at this extremely low metallicity is very  important since their stellar surface composition is not expected to be significantly modified by any internal mixing processes as in giant stars\citep{spi05}.
 J0815+4729 is similar to HE 1327--2326 in regards to its carbon enhancement, effective temperature and metallicity. 
HE 1327--2326 is considered a turn-off/subgiant star, while J0815+4729 appears to be a dwarf. The ISIS spectrum of HE 1327--2326 indicates a metallicity of [Fe/H]$\sim -4.9$, since stellar Ca line is blended in that 
spectrum with the ISM features~\citep{agu17I}.  However the authors proposed a simple analysis of the ISM effect based on the UVES spectrum of HE 1327--2326. For J0815+4729, we require a high-resolution spectrum 
to clearly isolate the stellar Ca feature from possible additional ISM lines, and thus, together with the 
detection of Fe lines, to establish the metallicity of this star. 
There are two other confirmed dwarf stars in this metallicity regime: one without any detectable carbon, J1029+1729~\citep{caff11}, and another carbon enhanced unevolved star, J1035+0641~\citep{boni15}. 
The  majority of extremely metal-poor stars show overabundances of carbon, [C/Fe]$>0.7$, and it appears that carbon enhanced metal-poor (CEMP) stars split into two groups, with dramatically different carbon abundances (see e.g. \citet{bee05,boni15,alle15} and references therein).
The two carbon-bands (high and low) studied have different origins. On the one hand, CEMP stars in the high-carbon band (A(C)$\sim8.2$) are probably produced by mass transfer from a binary companion, most likely an AGB star \citep{sta14}. On the other hand, objects lying in the low-carbon-band  (A(C)$\sim6.8$) are thought to show the original carbon  abundance inherited by the star from the interstellar medium \citep{stan09,aba16,boni15}. 
J0815+4729 has an abundance ratio of $\left[\rm C/\rm Fe \right]\geq+5.0$\,dex\, corresponding to A(C)$\sim7.7$\,dex (adopting [Fe/H]$\leq -5.8$). 
In Figure 4 (bottom panel) we show the carbon abundance ratio [C/Fe] for all stars below [Fe/H]$<-4.5$.  All stars in this metallicity regime are considered to belong to the low-carbon band \citep{boni15}, except for J0815+4729, which appears to be in between the low- and high-carbon bands.
 Both metallicity and carbon abundance are considered upper and lower limits, respectively.
 High-resolution spectra would be very useful to measure other elemental abundances and investigate the properties of the first supernovae.  In particular the barium abundance, or that of any other s-element is not measurable from ISIS or OSIRIS spectra, and this is required to determine whether J0815+4729 is a  CEMP-s, CEMP-r or i-process star \citep{ham16}. If we establish the abundance pattern we will learn about the progenitor properties. Finally, the radial velocity accuracy from medium-resolution data is not enough to discard variations 
among different exposures, which would be indicative of binarity. 

The discovery of J0815+4729 starts to fill in the gap in iron abundance between SMSS J01313--6708 and the rest of the extremely metal-poor stars. Identifying and characterizing chemically these rare breed of stars will certainly shed light on the early chemical evolution of the Galaxy and the nature of the first stars.


\begin{acknowledgements}
DA acknowledges the Spanish Ministry of Economy and Competitiveness 
(MINECO) for the financial support received in the form of a 
Severo-Ochoa PhD fellowship, within the Severo-Ochoa International PhD 
Program.
DA, CAP, JIGH, and RR also acknowledge the Spanish ministry project MINECO AYA2014-56359-P. JIGH acknowledges financial support from the Spanish Ministry of Economy and Competitiveness (MINECO) under the 2013 Ram\'on y Cajal program MINECO RYC-2013-14875. The authors thankfully acknowledge the technical expertise and assistance provided by the Spanish Supercomputing Network (Red Espanola de Supercomputaci\'on) and Antonio Dorta in particular, as well as the computer resources used: the LaPalma Supercomputer, located at the Instituto de Astrof\'isica de Canarias. This paper is based on observations made with the Gran Telescopio de Canarias (GTC) and with the William Herschel Telescope, operated by the Isaac Newton Group at the Observatorio del  Roque de los Muchachos, La Palma, Spain, of the Instituto de Astrof{\'i}sica de Canarias. We would like to thank GRANTECAN and ING staff members for their efficiency during the observing runs. In particular, we thank to A. Cabrera Lavers and D. Reverte Pay\'a for their help during OSIRIS observations.\\
\end{acknowledgements}
\bibliography{biblio}

\end{document}